\documentclass[12pt]{iopart}

\usepackage[english]{babel}
\usepackage{amsfonts}
\usepackage{amssymb}
\usepackage{amstext}
\usepackage{epsfig}

\def\openone{\leavevmode\hbox{\small1\kern-3.8pt\normalsize1}}
\newcommand{\beq}{\begin{equation}}
\newcommand{\eeq}{\end{equation}}
\newcommand{\barr}{\begin{eqnarray}}
\newcommand{\earr}{\end{eqnarray}}
\newcommand{\ket}[1]{\left\vert#1\right\rangle}
\newcommand{\Ham}{\mathcal H}
\newcommand{\Id}{{\cal I}}
\newcommand{\eps}{\varepsilon}
\newcommand{\HamEff}{{\mathbb H} }
\newcommand{\Norm}{{\mathbb N} }
\newcommand{\Morm}{{\mathbb M} }

\begin{document}
\title{Stiffness in 1D Matrix Product States with periodic boundary conditions}

\author{Davide Rossini, Vittorio Giovannetti and Rosario Fazio}

\address{Scuola Normale Superiore, NEST and Istituto Nanoscienze \, CNR, Pisa, Italy}

\date{\today}

\begin{abstract}

We discuss in details a modified variational matrix-product-state algorithm for 
periodic boundary conditions, based on a recent work by 
P. Pippan, S.R. White and H.G. Everts, Phys. Rev. B {\bf 81} (2010) 081103(R), 
which enables one to study large systems on a ring (composed of $N \sim 10^2$ sites).
In particular, we introduce a couple of improvements that allow to enhance
the algorithm in terms of stability and reliability.
We employ such method to compute the stiffness of one-dimensional 
strongly correlated quantum lattice systems.
The accuracy of our calculations is tested in the exactly solvable spin-1/2 
Heisenberg chain.

\end{abstract}


\section{Introduction}

The recent technological advances in manipulating cold atomic gases with very high
control and accuracy have opened up the possibility to test the quantum physics of
many-particle systems in its roots~\cite{bloch2008}. Moreover, quantum many-body 
systems can sustain collective states of matter which have no classical analog, 
such as superfluid or insulating quantum phases~\cite{greiner2002}, 
or even more exotic states such as the supersolid~\cite{kim2004}.
Experimental achievements along these direction renewed, in parallel, a considerable 
theoretical interest in the study of strongly correlated systems. 
Unfortunately, despite the apparent simplicity of their constituting Hamiltonian, 
the lack of a dominant exactly solvable 
contribution ultimately limits the applicability of conventional perturbative methods, 
thus drastically restricting analytic studies of both statics and non-equilibrium 
dynamical properties to very few cases. Approximated techniques or fully numerical 
approaches are therefore often required. 

In the early Nineties, White proposed a very powerful and accurate algorithm for numerically 
performing the renormalization group procedure in one-dimensional (1D) systems~\cite{white1992}.
Its large applicability to the study of both static and dynamic properties 
of the low-energy spectrum of generic strongly correlated 1D quantum systems, 
stimulated a considerable part of condensed matter theorists, which, 
in the subsequent years, produced a number of relevant results lying on this method, 
also called the Density Matrix Renormalization Group (DMRG) (see, e.g., 
Ref.~\cite{schollwockRMP} and references therein).
The formal equivalence of the DMRG procedure with a Matrix Product State 
(MPS)~\cite{ostlund1995}, which can be written in terms of a product of certain 
matrices, permitted a reformulation of the DMRG method as an optimization algorithm.
This allowed a better understanding of the renormalization procedure, together with
a number of promising extensions and improvements of the method, mostly coming 
from quantum information communities.
Nonetheless, unlike the original DMRG protocol, these new algorithms are still not of common 
use and their potentialities have not been completely exploited~\cite{schollwock2011}.

One of the drawbacks of the celebrated DMRG algorithm, in its canonical
formulation, is its intrinsic difficulty in simulating
ground states of 1D systems with periodic boundary conditions. 
In particular, the accuracy scales like the square of the number of states 
kept in an analogous simulation with open boundary conditions~\cite{white1992}.
For this reason, its applicability to systems closed on a ring is limited
to very few examples in the literature~\cite{scholl99}.
Much better performances can be obtained by reformulating the DMRG in terms of
a variational procedure on the class of MPS with periodic boundaries~\cite{verstraete2004}.
This is done at the expense of no longer using sparse matrices, thus reflecting in
a considerable slowdown of the procedure; moreover, due to the cyclic structure
of the MPS, contractions of the various matrices also become more costly.
However, from a conceptual point of view, the accuracy should drastically increase,
since the original DMRG was equivalent to an optimization method in which the variational 
class of states intrinsically belonged to the MPS with open boundaries.
Clever strategies for reducing the computational cost of such algorithms, 
in the case of translational invariant MPS, have been developed in 
Refs.~\cite{sandvik2007,shi2009,pirvu2010} and tested using Heisenberg and Ising 
spin-1/2 chains as benchmark models.
Very recently, an improved version of the algorithm in Ref.~\cite{verstraete2004},
for generic non-translationally invariant systems, has been put forward 
by Pippan {\it et al.}~\cite{pippan2010}. As we will outline in the following,
this enables a great computational speedup, so that precisions of the same 
order of magnitude of the ones obtained with open boundaries can be reached with 
a reasonable computational cost.

The purpose of this paper is twofold. First we present in detail 
an optimized variational MPS algorithm for periodic boundary conditions, 
which enables one to study static properties of the ground state of large, 
generically non-homogeneous, systems on a ring (made up of $N \sim 10^2$ sites).
The most important steps of the algorithm are those of Ref.~\cite{pippan2010},
we present however a number of suggestions that improve its stability.
Specifically, first we provide a  way to stabilize the generalized eigenvalue
problem that has to be solved at each variational step.
This consists in $\:$ {\em i)} suitably choosing the gauge conditions for the MPS, 
as previously hinted in Ref.~\cite{murg2008}, and then in $\:$ {\em ii)} wiping out the kernel
of the effective norm operator by introducing small perturbative corrections
in the effective operators. Furthermore, we also employ the decomposition of products of Ref.~\cite{pippan2010},
other than for multiplying transfer matrices, also to simplify the decomposition
of the effective Hamiltonian, thus speeding up its diagonalization.
The capabilities of the algorithm and the scaling of the precision with the
dimension of the MPS are the same as those in the standard DMRG algorithm.
Putting emphasis on all the major technical details of our algorithm, 
we aim at providing a rather comprehensive explanation which reveals itself 
compulsory for everybody who wants to implement it operatively.

The second purpose of the paper is to employ the algorithm in order to 
numerically evaluate the stiffness in one-dimensional models.
Its evaluation inherently requires periodic boundaries and high degree
of accuracy in the determination of energies, which can be reached
within our approach and with a relatively small computational effort.

The paper is organized as follows.
In Sec.~\ref{sec:algorithm} we describe our variational MPS algorithm valid for 
finding the ground state of 1D lattice systems on a closed ring,
discussing the strategies we adopted in order to make it reliable and efficient  (see Sec.~\ref{sec:stabilization}).
In Sec.~\ref{sec:Heisenberg} we provide a non trivial application of the scheme. 
In particular, we take advantage of the boundaries in order to compute the spin 
susceptibility in a 1D spin-$1/2$ Heisenberg chain, thus locating the critical phase 
of the system. Finally, in Sec.~\ref{sec:concl} we draw our conclusions
and outline some possible lines of investigation.

\section{MPS variational method for periodic boundary conditions} \label{sec:algorithm}

In this section we give a detailed overview of the numerical method employed in the simulations, 
with emphasis on the expedients we adopted in order to achieve the highest
possible degree of stability and accuracy, while keeping the computational cost
of the overall algorithm at minimum (see Sec.~\ref{sec:stabilization}).
\newline 

The ground state of a generic non translationally invariant one-dimensional (1D) 
quantum system on a lattice of $N$ sites
can be well approximated by a Matrix Product State (MPS) of the form:
\beq
\ket{\psi} = \sum_{i_1, i_2, \ldots , i_N} {\rm Tr} (A^{[1],i_1} \cdot A^{[2],i_2} \cdot \ldots  \cdot A^{[N],i_N} )
\ket{i_1}_1\ket{i_2}_2 \ldots \ket{i_N}_N \, ,
\label{eq:MPS}
\eeq
where $\ket{i_k}_k$ denote the $d$ basis states we selected to describe the $k$-th site of the system
(for the sake of clarity we take $d$ equal on all sites, without losing in generality), 
while $A^{[k],i_k}$ is a set of $d$ matrices of dimension $m \times m$ ($m$ is usually referred to as
the {\it bond link} of the MPS), and where  ``$\cdot$" represents the standard row-by-column 
matrix product (see, e.g., Ref.~\cite{cirac2009}).
The accuracy of such ansatz generally improves when increasing $m$ (as a matter of fact, any state 
of $N$ sites admits an explicit representation of the form~(\ref{eq:MPS}) with $m\sim d^{N/2}$). 
It turns out that, due to area law~\cite{eisert2010}, for ground states of generic 
1D systems, accurate descriptions (even in the critical regime, where area law is violated, 
but only through an additional logarithmic term in the system size) can be obtained 
with a bond link dimension $m$ which does not scale with the system length.
Such a representation therefore drastically reduces the amount of needed resources 
from exponential ($\sim d^N$) to polynomial growth ($\sim N d m^2$) with the system size $N$,
thus making simulations feasible.
It is also worth stressing that, for each $\ket{\psi}$, the representation~(\ref{eq:MPS}) 
is not unique: indeed the rhs of such expression is left invariant when replacing 
the $A^{[k],i_k}$ matrices according to 
\begin{eqnarray}
  A^{[k],i_k} \rightarrow A'^{[k],i_k}=
  V^{[k]}\cdot  A^{[k],i_k} \cdot W^{[k\oplus1]} \;,\label{gauge}
\end{eqnarray}
where  ``$\oplus$" stands for the sum modulus $N$, and where  
$\{ V^{[1]},V^{[2]}, \cdots, V^{[N]}\}$ and $\{ W^{[1]},W^{[2]}, \cdots, W^{[N]}\}$  
represent two sets of (non necessarily quadratic) matrices which 
fulfill the isometric condition $W^{[k]}\cdot V^{[k]}= {\cal I}$, with ${\cal I}$ being 
the $m\times m$ identity matrix. The freedom implied by Eq.~(\ref{gauge}) can be fixed by imposing 
proper {\em gauge} conditions to the matrix elements entering Eq.~(\ref{eq:MPS}) which, in some cases, 
happen to be helpful in speeding up the numerical search of the optimal ansatz state.

\subsection{Density Matrix Renormalization Group approach}

It has been proven that any real-space renormalization procedure results in 
an MPS structure~\cite{ostlund1995,verstraete2004},
therefore the best approximation to the ground state in an MPS-like form
of a given Hamiltonian $\Ham$ is obtained through a variational principle, 
that is by minimizing the energy function 
$E := {\cal E/N} = \langle \psi \vert \Ham \vert \psi \rangle / \langle \psi \vert \psi \rangle$ 
with respect to all the MPS of the type in Eq.~(\ref{eq:MPS}).

In the following we focus on 1D lattice Hamiltonians of finite size $N$, of the type:
\beq
\Ham = \sum_{k=1}^N \sum_{\alpha, \beta} h_k^{(\alpha,\beta)} \sigma^\alpha_k \otimes \sigma^\beta_{k+1} \, ,
\label{eq:HamGen}
\eeq
where $\sigma^\alpha_k$ denote some operator on site $k$ (e.g.,
the spin operators with $\alpha = x,y,z$ for a quantum spin lattice system),
while $h_k^{(\alpha,\beta)}$ are the coupling strengths.
Open Boundary Conditions (OBC) are set by imposing $\sigma^\alpha_{N+1} =0$,
while Periodic Boundary Conditions (PBC) can be fulfilled by requiring 
$\sigma^\alpha_{N+1} = \sigma^\alpha_1$.
The generalization to on-site and short-range interactions other than 
nearest-neighbor can be framed in the picture we are going to elucidate, 
even if in the following we will not explicitly discuss them.

For any operator $\sigma_k$ on site $k$, let us define the so called
{\it transfer matrix} $E^{[k]} [\sigma_k]$ of dimensions $m^2 \times m^2$:
\beq
  E^{[k]} [\sigma_k]  = \sum_{i,i'=1}^d \langle i' \vert \sigma_k \vert i \rangle \, (A^{[k],i'} )^* \otimes A^{[k],i} \, ,
\eeq
where $(\cdot)^*$ denotes the complex conjugate. 
In this way we can write the expectation value ${\cal E} = \langle \psi | {\cal H} |\psi\rangle$ 
of Hamiltonian~(\ref{eq:HamGen}) on a generic MPS state~(\ref{eq:MPS}) as
\begin{eqnarray}
  {\cal E} =  \sum_{k=1}^N \sum_{\alpha, \beta} h_k^{(\alpha,\beta)} \; 
  \langle E^{[1]} \cdot \ldots 
  \cdot E^{[k]} [\sigma^\alpha_k] \cdot E^{[k+1]} [\sigma^\beta_{k+1}] \cdot E^{[k+2]} \cdot \ldots \cdot E^{[N]} \rangle
  \label{eq:energy}
\end{eqnarray} 
where we adopted the abbreviations $E^{[k]} [\openone_k] := E^{[k]}$ to describe the transfer matrix 
of the identity operator $\openone_k$, and $\langle \cdots \rangle := \mbox{Tr}[ \cdots ]$
to represent the trace operation. 
A similar expression holds for the norm ${\cal N}=\langle \psi | \psi\rangle$ of the MPS, which is written
as a simple product of transfer matrices $E^{[k]}$ on the identity matrix, i.e.,
${\cal N} = \langle E^{[1]} \cdot E^{[2]} \cdot \ldots \cdot E^{[N]} \rangle$.

This equation elucidates the dependence of the energy ${\mathcal E}$ on the matrices $A^{[k],i}$,
and in principle it can be used to determine them following a variational technique that
minimizes ${\cal E}$.
For generic non translationally invariant systems, as we are supposing from the beginning,
such a minimization problem would contain a huge number of parameters, that is of the order
$\sim O(N d m^2)$, and would be operatively intractable. 
In practice, standard optimization techniques, like the Density Matrix Renormalization
Group (DMRG) approach, adopt clever schemes to achieve the minimization.
The key point of the algorithm is the following: one sequentially minimizes
the energy with respect to the $A$'s by fixing all of them except the ones on a given site $k$
(rigorously speaking, the standard celebrated DMRG algorithm optimizes two sites at
the same time~\cite{white1992}.
As a matter of fact, in particularly adverse cases single-site optimizations may get
stuck in local minima or converge much slowly, due to the suppression of fluctuations 
between the system and the environment~\cite{white2005}).
As it is apparent from Eq.~(\ref{eq:energy}), and the analogous equation for ${\cal N}$,
the dependence on $A^{[k],i}$ is only quadratic.
The minimization of energy thus consists of minimizing a quadratic polynomial ${\cal E}$
with quadratic constraints ${\cal N}$, which operatively translates into solving
a generalized eigenvalue equation of the form:
\beq
\HamEff_k x_k = \lambda \Norm_k x_k \,.
\label{eq:GenEV}
\eeq
Here we have formally mapped the unknown coefficients of the $m \times m$ matrices
$A^{[k],i}$  (with $i=1,\ldots,d$),
on a column vector $x_k$ of dimensions $d m^2$, while the $dm^2\times d m^2$ matrices $\HamEff_k$
and $\Norm_k$ (which hereafter will be referred to respectively as the
``effective Hamiltonian'' and the ``effective norm operator'')
can be straightforwardly determined by employing the same mapping
on Eq.~(\ref{eq:energy}) and analogous for ${\mathcal N}$. 
In particular we notice that $\Norm_k$ can be always written as a tensor product 
of a $m^2\times m^2$ matrix $\Morm_k$ times a $d\times d$ identity matrix $I$ 
associated with the index $i$ of $A^{[k],i}$~\cite{murg2008}, i.e.,
\beq
  \Norm_k := \Morm_k \otimes I \;. \label{MORM} 
\eeq
It is also worth stressing that, by calling $\ket{\psi(x_k)}$ the many body 
MPS state~(\ref{eq:MPS}) obtained by contracting matrix $A^{[k],i}$ 
corresponding to the vector $x_k$ with all the other matrices at sites different from $k$,
the matrices $\HamEff_k, \Norm_k$ satisfy the following identities,
\beq
  \langle \psi(y_k) | {\cal H} | \psi(x_k)\rangle = y_k^\dag \; \HamEff_k \; x_k \;, \qquad 
  \langle \psi(y_k)  | \psi(x_k)\rangle = y_k^\dag \; \Norm_k \; x_k \;,  \label{NEWNEWEQ1}
\eeq
for all choices of the column vectors $x_k$ and $y_k$. Accordingly it follows that
$\HamEff_k, \Norm_k$ are Hermitian, that $\Norm_k$ is non negative, and that 
the kernel of $\Norm_k$ (if any) is always included in the kernel of $\HamEff_k$,
i.e., $\mbox{kern}(\Norm_k) \subseteq \mbox{kern}(\HamEff_k)$.
The last inclusion follows from the fact that, if $x_k \in \mbox{kern}(\Norm_k)$, 
then the associated many body vector state $|\psi(x_k)\rangle$ is identically null 
(indeed, according to the second expression of Eq.~(\ref{NEWNEWEQ1}), it has null norm). 
Therefore, from the first identity of Eq.~(\ref{NEWNEWEQ1}) we have that 
$y_k^\dag \; \HamEff_k \; x_k = \langle \psi(y_k) | {\cal H} | \psi(x_k)\rangle  =0$ 
for each column vector $y_k$. This implies that $\HamEff_k \; x_k$ nullifies, 
proving that $x_k$ is indeed an element in the kernel of $\HamEff_k$.

Once the matrices $A^{[k],i}$ associated with the $k$-th site have been optimized, 
the next step consists in minimizing
with respect to $A^{[k+1],i}$ and so on, until the rightmost site $N$ has been reached.
One then proceeds analogously going leftward from site $N$ to site $1$, that is,
one sweeps through the spins from left to right and vice-versa, determining at
each step the matrices associated with a particular site.
This variational procedure eventually converges to a minimum of the energy\footnote{
It is worth noticing however that, as common in these numerical optimization strategies, 
there is no formal proof that the reached minimum  will be an absolute one 
(indeed it is possible that it will be only a local minimum). Of course one could improve 
the confidence of the procedure by performing optimization iterations 
that work on the optimization of the tensors pertaining to (say) a couple of consecutive sites.}.
It turns out that, for {\it open boundary conditions} (OBC), the problem can be considerably simplified
by exploiting the freedom~(\ref{gauge}) to impose suitable gauge conditions on the matrices $A^{[k],i}$ 
so that, at each step, the state $\ket{\psi}$ is always normalized {\em and} $\Norm_k$ 
can be kept equal to the identity matrix. 
Equation~(\ref{eq:GenEV}) would then become a standard eigenvalue equation of the type:
\beq
\HamEff_k x_k = \lambda x_k \,.
\label{eq:StdEV}
\eeq
To verify this, we first notice that since the first and the last matrix entering the MPS 
expression~(\ref{eq:MPS}) need not to be directly connected via a dedicated bound link, 
one can assume they are expressible as $\left[A^{[1],i}\right]_{\ell, \ell'} = \delta_{\ell,1}\;  a^{[1],i}_{\ell'}$, 
$\left[A^{[N],i}\right]_{\ell, \ell'} = a^{[N],i}_{\ell}\; \delta_{\ell',1}$, 
with $\delta_{\ell,\ell'}$ being the Kronecker delta.
The gauge conditions for the optimization on site $k$ consist then in choosing 
a so called ``left isometry condition'' for all the matrices on the left of~$k$:
\beq
\sum_{i=1}^d (A^{[j],i})^\dagger \cdot A^{[j],i} = \Id \;,\qquad \forall j = 1, \ldots, k-1 \,,
\label{eq:LeftIso}
\eeq
and a ``right isometry condition'' for matrices on the right of $k$:
\beq
\sum_{i=1}^d A^{[j],i} \cdot (A^{[j],i})^\dagger = \Id \qquad \forall j = k+1, \ldots, N \,,
\label{eq:RightIso}
\eeq
with ${\cal I}$ being the $m\times m$ identity matrix. 
In this way ${\cal N}$ becomes a constant (i.e., ${\cal N} = m$) which can be trivially 
absorbed in the definition of $\HamEff_k$.

Going from left to right, condition~(\ref{eq:LeftIso}) on the optimized matrices $A^{[k],i}$ 
is enforced by performing a Singular Value Decomposition (SVD) on the matrix 
$V_{i\alpha, \beta} = A^{[k],i}_{\alpha, \beta}$ ($\alpha, \beta = 1, \ldots, m$
are row and column indices of the matrix $A^{[k],i}$; the site index $i$ has been grouped 
into the row index of the matrix $V$ -- in this way Eq.~(\ref{eq:LeftIso}) equals to say 
$V^\dagger V = {\cal I}$) such to obtain $V=UDW$ with $U,W$ isometries
and $D \geq 0$ diagonal matrix. One then discards $DW$ and simply substitutes the $A^{[k],i}$
matrices with $A'^{[k],i}_{\alpha, \beta} = U_{i\alpha, \beta}$.
Going from right to left, one does a similar procedure after grouping site index $i$
into the columns of the $V$ matrix: $V_{\alpha, i \beta} = A^{[k],i}_{\alpha, \beta}$
(so that Eq.~(\ref{eq:LeftIso}) translates into $V V^\dagger = {\cal I}$),
and performing a SVD on $V$.
The matrices $A^{[k],i}$ are then changed with $A'^{[k],i}_{\alpha, \beta} = W_{\alpha, i \beta}$.

\subsection{Constructing the effective Hamiltonian}

The core of the DMRG algorithm consists in the iterative resolution of Eq.~(\ref{eq:StdEV})
in the case of OBC or of Eq.~(\ref{eq:GenEV}) for PBC.
Nonetheless, even the construction of the effective Hamiltonian $\HamEff_k$ 
(and of the effective norm operator $\Norm_k$ for PBC) may constitute a bottleneck.
As a matter of fact, it turns out that, for OBC, the number of operations that are
needed to construct $\HamEff_k$ is relatively small and scales as $m^3$.

In order to achieve such a goal, at every step of the variational algorithm 
it is convenient to store some operators that are products of transfer matrices.
Let us suppose we are optimizing the $k$-th site. In this case we may want to save the matrices:
\beq
\begin{array}{rl}
{\mathcal T}^{[<m]} := & E^{[1]} \cdot E^{[2]} \cdot \ldots \cdot E^{[m-1]}\;, \\
{\mathcal B}_\alpha^{[<m]} := & E^{[1]}\cdot E^{[2]} \cdot \ldots \cdot E^{[m-1]}[\sigma^\alpha_{m-1}]\;, \\
{\mathcal H}^{[<m]} : = & \sum_{j=1}^{m-2} \;h_j^{(\alpha,\beta)}  \; E^{[1]}\cdot \ldots 
\cdot E^{[j]}[\sigma^\alpha_j] \cdot E^{[j+1]}[\sigma^\beta_{j+1}] \cdot \ldots\cdot E^{[m-1]} \;,\\
\end{array}
\label{eq:Eproducts}
\eeq
for each $m<k$ and analogous ones for every $m>k$.
In this way one has
\barr
    {\mathcal E} & = & \langle  {\mathcal H}^{[<k]} \cdot E^{[k]} \cdot {\mathcal T}^{[>k]} \rangle +
    \langle {\mathcal T}^{[<k]} \cdot E^{[k]} \cdot {\mathcal H}^{[>k]} \rangle   \label{eq:HamBuildup}  \\ 
    {} & {} & + \sum_{\alpha, \beta}  \; h_k^{(\alpha,\beta)} \; \langle 
    {\mathcal T}^{[<k]} \cdot E^{[k]}[\sigma^\alpha_k] \cdot {\mathcal B}_\beta^{[>k]} + 
    {\mathcal B}_\alpha^{[<k]} \cdot E^{[k]}[\sigma^\beta_k] \cdot {\mathcal T}^{[>k]} \; \rangle  \, , \nonumber \\
    {\mathcal N} & = & \langle {\mathcal T}^{[<k]} \cdot E^{[k]} \cdot {\mathcal T}^{[>k]} \rangle  \;, 
    \label{eq:NormBuildup} 
\earr
so that the effective Hamiltonian $\HamEff_k$ and the norm operator $\Norm_k$ 
can be explicitly constructed. 
Quite remarkably, one can see that, when going to the next iterative step at site $k+1$,
the corresponding matrices ${\mathcal T}^{[<k+1]}, {\mathcal B}_\alpha^{[<k+1]}, {\mathcal H}^{[<k+1]}$ 
can be built up from the corresponding ones with indexes  ${}^{[<k]}$ by multiplying them 
by the  transfer matrix corresponding to site $k$ and by properly adding the result. 
In the case of OBC, each one of these passages requires one to perform a number of fundamental operations 
which scales as $m^3$.
Besides that, since the $A$ matrices of Eq.~(\ref{eq:MPS}) are chosen such to fulfill an isometry condition
of the type~(\ref{eq:LeftIso}) (left of $k$) or~(\ref{eq:RightIso}) (right of $k$), 
the operators ${\mathcal T}$ are trivial and correspond to the identity.
This implies that the effective Hamiltonian~(\ref{eq:HamBuildup}) is a sparse matrix, thus dramatically speeding up 
the resolution of the eigenvalue problem in Eq.~(\ref{eq:StdEV})~\footnote{
Since only the eigenstate corresponding to the smallest eigenvalue is needed,
powerful numerical methods such as Davidson or Lanczos techniques can be suitably used.
Unlike brute-force diagonalization approaches, these methods also take advantage 
of the sparseness of the matrix, requiring only a matrix-vector multiplication routine.}.

With PBC, two major obstacles emerge.
On one hand, the resolution of a generalized eigenvalue problem of the type in Eq.~(\ref{eq:GenEV}) 
can have problems if the matrix $\Norm_k$ is ill-conditioned, while one in general 
can no longer benefit of the sparseness of matrix $\HamEff_k$. 
On the other hand, the number of operations required to build up $\HamEff_k$
and $\Norm_k$ is considerably larger, due to the absence of boundaries from
which performing the contractions.
It turns out that each of the basic contractions that are needed,
that is the multiplication of a transfer matrix of dimensions $m^2 \times m^2$
with a product of transfer matrices of the same dimensions, require $O(m^5)$ operations.
This drastically limits the capabilities of the algorithm to very small sizes ($N \sim 20 \div 30$)
with a low bond link~\cite{verstraete2004}.

\subsection{Truncated SVD} \label{sec:SVD}

Remarkably, as discussed in Ref.~\cite{pippan2010}, from a computational point of view
the second obstacle can be overcome by using the following observation: 
if one performs a SVD decomposition of a sufficiently long 
product of $m^2 \times m^2$ transfer matrices ($k\gg1$), 
the singular values $\Sigma_j$ in general will decay fast, i.e.: 
\beq
E^{[1]} \cdot E^{[2]} \cdot \ldots \cdot E^{[k]} \approx \sum_{j=1}^p \Sigma_j \, {\mathbb U}_j \, {\mathbb V}_j^T \;,
\label{eq:SVDtrunc}
\eeq
where $p \ll m^2$ ($m^2$ being the total number of singular values of the product in the lhs),
while ${\mathbb U}_j$ and ${\mathbb V}_j$ respectively denote a left and a right
singular vector of size $m^2$.
Intuitively Eq.~(\ref{eq:SVDtrunc}) can be justified by the fact that, in the limit of large $N$, 
the local physics of the system should not really be affected by the properties of the boundaries: 
consequently considering that, for OBC, imposing the gauge conditions~(\ref{eq:LeftIso}),
(\ref{eq:RightIso}) is formally equivalent to enforce exactly the structure of Eq.~(\ref{eq:SVDtrunc}) 
with $p=1$, one thus expects that, for PBC, the rhs of Eq.~(\ref{eq:SVDtrunc}) 
should constitute a good approximation of the lhs.

Therefore, if one knows a priori that $p \ll m^2$, he can evaluate products of transfer matrices
of the type in Eq.~(\ref{eq:Eproducts}) by performing a truncated SVD which enables
the calculation of only the largest $p$ singular values out of the $m^2$ possible outcomes.
Such an operation requires only $O(p \times m^3)$, that is a $m^2$ factor less than
the standard SVD. However we verified that, in order to achieve good accuracies 
for the sizes we have considered, in practice one has to choose a value of $p$ 
which scales approximately linearly with $m$: $p \sim m$. This limits the actual gain 
of the overall operation by a factor of $m$.~\footnote{
In Ref.~\cite{pippan2010} a gain of a factor $m^2$ over the plain algorithm is claimed, 
with the truncated SVD strategy. We point out that, within our numerical experience,
we could reach adequate precisions for our measures only by scaling $p$ linearly with $m$
(see, e.g., data in Fig.~\ref{fig:Energy_Heis}).
This in practice would reduce the computational effort only from $m^5$ to $m^4$.}
The method is explained below in the Sec.~\ref{subsec:SVD}.
Once the product of $k$ transfer matrices has been written in the form of Eq.~(\ref{eq:SVDtrunc}),
multiplying it with another matrix $E^{[k]}$ on the right of it such
to obtain ${\mathcal T}^{[>k+1]}$ is relatively easy and also requires $O(p \times m^3)$ operations,
involving only the $p$ vectors ${\mathbb V}_j^T$ [similarly, multiplying it on the left by
a transfer matrix requires acting only on ${\mathbb U}_j$ with $O(p \times m^3)$ operations].

\subsubsection{Operative algorithm for the truncated SVD} \label{subsec:SVD}

The SVD of a generic matrix $M \in {\cal M}_{m^2 \times m^2}$ generally requires $O(m^6)$ 
elementary operations [more precisely, taking into account the tensorial product structure 
of the transfer matrices, in this specific case $O(m^5)$ operations are needed].
If one is only interested in the contribution coming from the $p$ largest singular values of $M$,
one can employ a truncated SVD according to the following prescription (see, e.g., Ref.~\cite{pippan2010} 
and~\ref{appa} for details).
Taking advantage of the tensor product structure of the transfer matrices, this 
requires only $O(p \times m^3)$ operations. 

$\bullet$ Generate a random matrix $x \in {\cal M}_{p \times m^2}$ of full rank;

$\bullet$ Multiply $x$ with the input matrix $M$ on the right: $y = x M$;

$\bullet$ Orthonormalize the rows of $y$ by using a Gram-Schmidt decomposition, so to
construct a matrix $y' \in {\cal M}_{p \times m^2}$;

$\bullet$ Take the transpose conjugate of $y'$ and multiply it with the input matrix $M$ 
on the left: $z = M (y')^\dagger$;

$\bullet$ Perform a SVD of such obtained $z \in {\cal M}_{m^2 \times p}$ matrix,
and write it as $z = UDV'$;

$\bullet$ Evaluate $V = V' y'$, so that $M$ can be written in SVD form as $M = UDV$.

\subsection{Stabilization of the generalized eigenvalue problem} \label{sec:stabilization}

Let us now come to the core of the optimization procedure, once the effective Hamiltonian 
and the effective norm are built up. We will discuss this point by presenting
few ``tricks" that we found useful to implement, in order to enhance the stability of the algorithm.

As it has been mentioned before, the generalized eigenvalue problem in Eq.~(\ref{eq:GenEV}) 
is typically harder to solve than the standard one~(\ref{eq:StdEV}).
Firstly, from a technical point of view, when using a periodic structure for the MPS,
it involves non-sparse matrices $\HamEff_k$ and $\Norm_k$.~\footnote{
Fast methods for finding the low-energy spectrum of large matrices, like the widely
used Davidson or Lanczos techniques, typically require to provide the application 
of the effective Hamiltonian and norm operator onto a generic input vector $\xi$.
With OBC this enables a great computational speedup due to the sparseness of the matrices;
with PBC such speedup unfortunately vanishes.}
This is because of the lack of a starting point (the two outermost sites 
of the chain, for OBC) from which the left~(\ref{eq:LeftIso}) and right~(\ref{eq:RightIso}) 
isometry conditions could be recursively applied.
As a consequence, with PBC the pure transfer matrices ${\mathcal T}$ can no longer 
be written as identities, and operators~(\ref{eq:HamBuildup})-(\ref{eq:NormBuildup})
are generally mapped into two non-sparse $d m^2 \times d m^2$ matrices, thus
inevitably slowing down the efficiency of the diagonalization procedure.
A second point, more of conceptual nature, is related to the conditionability of the problem.
Equation~(\ref{eq:StdEV}) is well conditioned, provided the spectrum of matrix $\HamEff_k$ is bounded.
This is not sufficient for Eq.~(\ref{eq:GenEV}), where one also necessarily has 
to ensure that the matrix $\Norm_k$ is strictly positive. 
The emergence of non trivial kernel spaces for $\Norm_k$ indeed introduces a critical 
instability in the diagonalization algorithms, which are usually based on convergence criterions 
of the type $\| \HamEff_k \, \xi_k - \lambda \xi_k \| / \| \Norm_k \, \xi_k \| < \mu$,
with $\mu$ small parameter controlling the convergence to the solution
(typical values are $\mu \lesssim 10^{-10}$).

To cope with this problem, we found it convenient to wipe out the kernel of $\Norm_k$ by forcedly adding
a small correction $\Norm_k \to \Norm_k(\eps) := \Norm_k + \eps \, \Id$ and 
$\HamEff_k \to \HamEff_k(\eps):= \HamEff_k + \sqrt{\eps} \, \Id$. 
Indeed, considering that the kernel of $\Norm_k$ is included into the kernel of $\HamEff_k$, 
it follows that the generalized eigenvalues associated with the matrices $\HamEff_k(\eps)$, 
$\Norm_k(\eps)$ form two sets: the first is composed by elements 
$\lambda(\varepsilon) \simeq \lambda +  O(\sqrt{\eps})$, with $\lambda$ being the generalized 
eigenvalues of the couple $\HamEff_k, \Norm_k$; the second instead is formed by terms which scale 
as $1/\sqrt{\varepsilon}$. Therefore, for $\varepsilon\rightarrow 0$,  
the minimum value of the $\lambda$ can be computed 
as the minimum generalized eigenvalue of $\HamEff_k(\eps)$, $\Norm_k(\eps)$ -- the only price to pay 
is the fictitious introduction of an error of order $\sim O(\sqrt{\eps})$.  
For practical purposes, reasonable values in order to remove instabilities, without substantially
affecting the simulation outcomes, are $\varepsilon \lesssim 10^{-12}$.

Once the kernel of $\Norm_k$ has been eliminated, other non-critical instabilities 
due to numerical accuracy may emerge
if eigenspaces of very small (positive) eigenvalues are present.
As pointed out in Refs.~\cite{verstraete2004,murg2008}, it is generally
very helpful, for an accurate numerical convergence of the code, to take advantage of the
gauge arbitrariness and redefine the matrices $\HamEff_k$ and $\Norm_k$ in such a way as to
increase the smallest eigenvalue of $\Norm_k$ as much as possible.
To this aim it is useful first to perform an approximate SVD on the non trivial 
$m^2 \times m^2$ component of ${\Norm}_k$, i.e. the matrix $\Morm_k$ of Eq.~(\ref{MORM}).
Using the recipe of Sec.~\ref{subsec:SVD}, we thus write 
\beq
  \Morm_k \approx \sum_j \Sigma^{\Morm_k}_j \;\;  {\mathbb U}^{\Morm_k}_j \;({\mathbb V}^{\Morm_k}_j)^T \, ,
   \label{eq:svdN}
\eeq
where $\Sigma^{\Morm_k}_1, \Sigma^{\Morm_k}_2, \cdots,$ are the singular values of the matrix $\Morm_k$ in decreasing order.
Here we remark that, if we used OBC, at any point $k$, the left and the right side 
of the lattice would have become factorized so that $\Morm_k$ was a tensor product
[besides, using suitable gauges expressed by the left and right isometry 
conditions~(\ref{eq:LeftIso})-(\ref{eq:RightIso}), it would have also been possible 
to write it as a tensor product of identities].
On the contrary, PBC originate unavoidable correlations between the two sides.
Nonetheless, for sufficiently long systems, these correlations can be reasonably considered
small, so that $\Morm_k$ is effectively close to a tensor product.
This implies that, in Eq.~(\ref{eq:svdN}), the largest singular value $\Sigma^{\Morm_k}_1$ 
is by far greater than all the others.

A clever gauge on the MPS structure can now be imposed such that the leading 
factorized term associated with the decomposition of Eq.~(\ref{eq:svdN}) 
gets close to the identity operator. It can be constructed by taking the $m \times m$ matrices 
$[N_1]_{\alpha, \beta} := (\Sigma^{\Morm_k}_1)^{-1/2} \; [{\mathbb U}^{\Morm_k}_1]_{\alpha, \beta}$ and 
$[N_2]_{\gamma, \delta} := (\Sigma^{\Morm_k}_1)^{-1/2} \;   [{\mathbb V}^{\Morm_k}_1]_{\gamma, \delta}$, 
where $\Sigma^{\Morm_k}_1, {\mathbb U}^{\Morm_k}_1$, and ${\mathbb V}^{\Morm_k}_1$ 
define the leading contribution of Eq.~(\ref{eq:svdN}).
The $m \times m$ (invertible) gauge matrix $X$ for the left side of the system can then be
constructed by solving the equation $X N_1 X^\dagger = \Id$
(an analogous matrix $Y$, which satisfies $Y N_2 Y^\dagger = \Id$,
is built up for the right side)~\footnote{
This equation can be formally solved by noting that $N_1 = X^{-1} (X^\dagger)^{-1} =
(X^\dagger X)^{-1}$ so that $X^\dagger X = N_1^{-1}$ and therefore $X = \sqrt{N_1^{-1}}$.
A solution is then obtained by taking the inverse square root of $N_1$, if this last
matrix is invertible. In case of a non-null kernel of $N_1$, the Moore-Penrose pseudoinverse 
$\sqrt{\tilde{N_1}^{-1}}$ can be considered.}.
%
\begin{figure}
  \begin{center}
    \vspace*{2mm}
    \includegraphics[width=13cm]{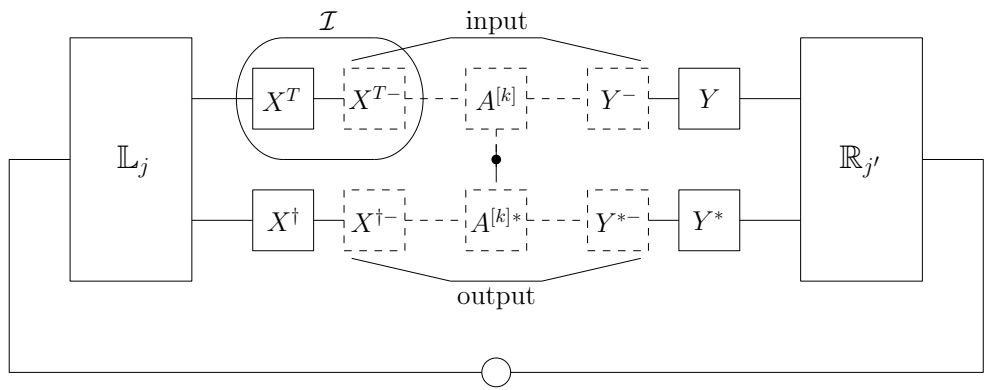}
    \caption{Stabilization scheme for the generalized eigenvalue problem.
      A suitable gauge is provided by the invertible matrices $X$ ($Y$)
      (the indexes ${}^{-}$ indicate inverse matrices, so that product matrices
      of the type highlighted in the oval blob exactly equal the identity matrix).
      All the left and right operators (${\mathbb L}_j$ and ${\mathbb R}_{j'}$) that are
      needed to build up $\HamEff_k$ and $\Norm_k$ [see Eqs.~(\ref{eq:HamBuildup})
      and~(\ref{eq:NormBuildup}) respectively] have to be contracted and transformed 
      according to Eqs.~(\ref{eq:HamEffGauge})-(\ref{eq:NormGauge}).
      Consequently, the input/output eigenvectors (dashed lines and boxes) are transformed 
      following Eq.~(\ref{eq:PsiGauge}).
      The black dot connecting input and output matrices stands for the 
      physical operators of the Hamiltonian at site $k$ that have to be accounted for,
      in some terms of Eq.~(\ref{eq:HamBuildup}).}
    \label{fig:stabilization}
  \end{center}
\end{figure}
%
Such gauge maps the eigenvalue equation~(\ref{eq:GenEV}) into
\beq
   \HamEff'_k x'_k = \lambda \Norm'_k x'_k \,,
   \label{eq:GenEVGauge}
\eeq
with
\barr
   \HamEff'_k = (X \otimes Y) \, \HamEff_k \, (X^\dagger \otimes Y^\dagger)  \;,\label{eq:HamEffGauge} \\
   \Norm'_k   = (X \otimes Y) \, \Norm_k   \, (X^\dagger \otimes Y^\dagger) \;,  \label{eq:NormGauge}   \\
   x'_k       = ( [X^\dagger]^{-1} \otimes [Y^\dagger]^{-1} ) \, x_k \label{eq:PsiGauge} \, .
\earr
As it is apparent from Eq.~(\ref{eq:NormGauge}), the new effective norm operator $\Norm'_k$ 
satisfies the property we required at the beginning, since the leading term 
has been transformed into the identity operator~\footnote{We remark that 
the gauge transformation presented here is not related with the one 
used in Ref.~\cite{pippan2010}. As a matter of fact, in our simulations we implemented 
both of them, but found that stabilization is better achieved with the one discussed in the text.}.
The application of the gauge, mapping the original eigenvalue problem~(\ref{eq:GenEV})
into~(\ref{eq:GenEVGauge}), can be operatively implemented by multiplying all the 
left(right)-side operators that are needed in order to build up $\HamEff_k$ and $\Norm_k$
[for an explicit expression, see Eqs.~(\ref{eq:HamBuildup})] by matrix $X$ ($Y$), 
as graphically shown in Fig.~\ref{fig:stabilization}.
Once Eq.~(\ref{eq:GenEVGauge}) is solved, one has to keep in mind that also the solution
$x'_k$ is gauged with the same matrices.

Finally we notice that the resolution of Eq.~(\ref{eq:GenEVGauge}) can be considerably speeded up 
by approximating ${\mathbb H}_k'$ along the line of what already done in Sec.~\ref{subsec:SVD}. 
Explicitly, we expanded ${\mathbb H}_k'$ via SVD and kept only the contributions 
associated with its $s$ largest eigenvalues [typically $s$ can be kept of the same order 
of the parameter $p$ of Eq.~(\ref{eq:SVDtrunc})].
By doing so, the number of fundamental operations needed to solve 
Eq.~(\ref{eq:GenEVGauge}) through standard large-matrix eigenvalues solvers, such as
the Lanczos method, can be drastically reduced.
In practice, by operating a cut in the SVD representation of ${\mathbb H}_k'$, 
one substantially improves the efficiency of the matrix-vector multiplication routine: 
$y_k' = {\mathbb H}_k' x_k'$ [where $x_k', \, y_k'$ are generic input/output $(d \times m^2)$-dimensional vectors].
This is crucial, since for matrix dimensions of the order of the ones considered in our MPS simulations,
the routine is typically called $O(10^2 \div 10^3)$ times,
and therefore its repeated iteration constitutes the actual bottleneck 
of the resolution of Eq.~(\ref{eq:GenEVGauge}).

More in the specific, we have to stress that the scaling of the contractions needed to perform 
the Hamiltonian-vector multiplication is not substantially modified, 
going from $O(p \times d \times m^3)$ to $O( s \times d \times m^3)$.
Nonetheless, what really changes is the prefactor of these scalings. 
This can be quite large ($\sim 20$) in the first case, accounting for the fact that
the effective Hamiltonian is made up of various terms coming from the fictitious division in sectors
of the system (see Sec.~\ref{sec:opt} for details) and their interconnections,
and also from the requirement of site $k$ to be kept free in order to employ the variational optimization.
On the other side, in our procedure the prefactor basically equals the unit, being it 
the fastest way to perform the multiplication, while essentially keeping the same Hamiltonian structure.
We carefully checked that this procedure does not notably affect the accuracy of the operation.

\subsection{Optimization algorithm} \label{sec:opt}

\begin{figure}
  \begin{center}
    \includegraphics[width=13cm]{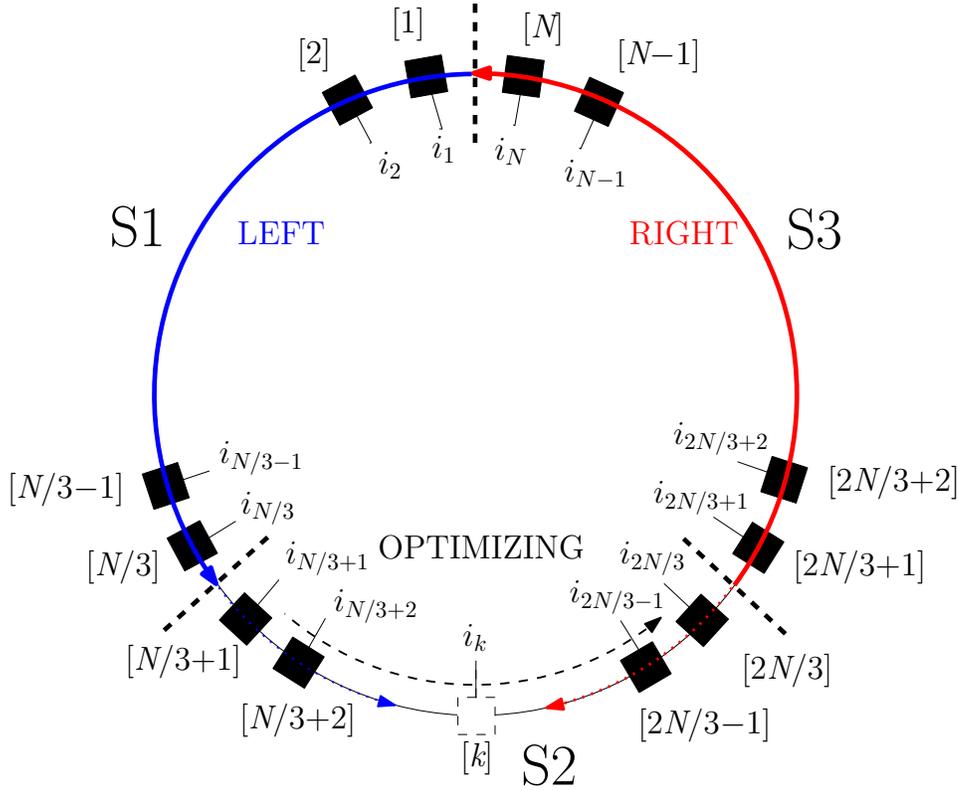}
    \caption{Circular optimization algorithm: the lattice ring of length $N$ is divided 
      into three sectors (S).
      The single optimization sweep proceeds from site $1$ to site $N$.
      When a new sector is entered (S2 in the example given in the figure), one builds up the
      product of transfer matrices for the other two sectors (S1 and S3).
      At the $k$-th optimization step, before performing the minimization in terms
      of the generalized eigenvalue problem~(\ref{eq:GenEV}), in order to build up the
      effective matrices $\HamEff_k$ and $\Norm_k$, one contracts the transfer matrices in the optimizing 
      sector (S2) to the left (right) of the optimizing site with the left (right) truncated SVD,
      following the order given by the arrows.}
    \label{fig:circular}
  \end{center}
\end{figure}

A single-site minimization procedure which is very convenient for 1D systems with PBC
is the circular scheme proposed in Ref.~\cite{pippan2010}, and here depicted in Fig.~\ref{fig:circular}.
Basically, the optimization proceeds following a circular pattern, rather than using
the standard backward and forward pattern that is generally used for OBC.

In order to guarantee that the number of transfer matrices that have to be multiplied
such to form Eq.~(\ref{eq:SVDtrunc}) is sufficiently large, one divides the circular
ring into three concatenated sectors of spins (reasonable values for the global
system size are $N \gtrsim 10^2$).
Optimization then starts from one section, say the first one, and proceeds 
along a fixed direction, say counterclockwise from site $1$ to site $N/3$.
Before initiating it, one has to construct the partial effective Hamiltonian 
and all the required operators corresponding to the two other sectors
(second sector from site $N/3+1$ to site $2N/3$, and third sector from site $2N/3+1$
to site $N$) in a SVD-like fashion, following the prescription of Sec.~\ref{subsec:SVD}.
Then a set of operators for the optimizing section can be constructed by
successively adding transfer matrices from the rightmost part of the optimizing
section to the left of the operators constructed for the section on the right.
After this, the normal optimization procedure can go on, until the border of the
optimizing section has been reached, say site $N/3$.
At that point, the procedure repeats from the beginning, with the section on the right
(from site $N/3+1$ to site $2N/3$) as the new optimizing section. 
And so on, in a circular way. A pictorial representation of the scheme discussed here
is given in Fig.~\ref{fig:circular}.

\begin{figure}
  \begin{center}
    \vspace*{5mm}
    \includegraphics[width=14cm]{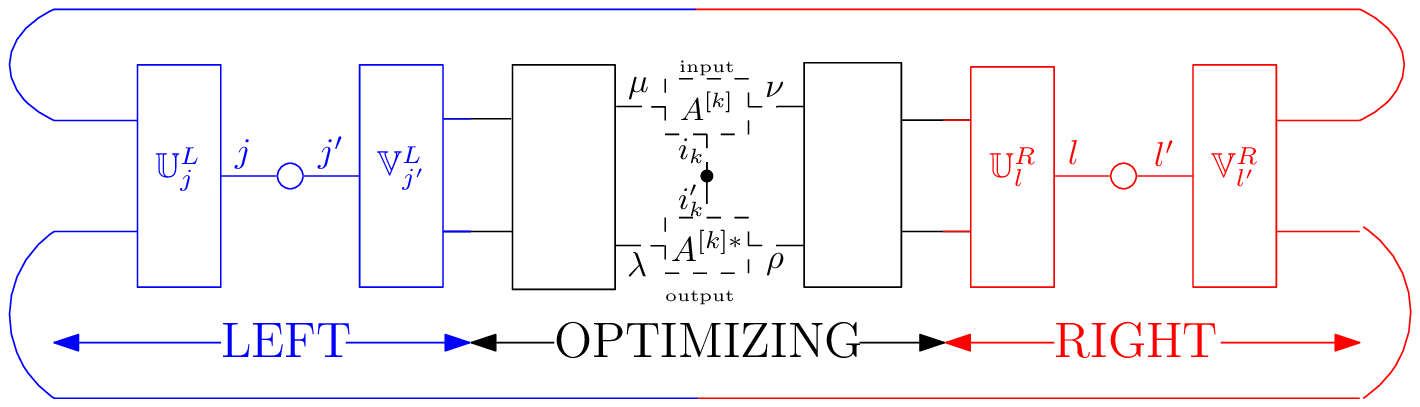}
    \caption{At each optimization step, the matrices $\HamEff_k$ and $\Norm_k$ are built up
      by contracting the transfer matrices of the three sectors (each line connecting
      the various blocks indicates a contraction).
      Due to the truncation in the SVD of the left and right sectors, links $j, j'$
      and $l, l'$ are made up of only $p$ elements. This makes it possible
      to considerably reduce the computational time for the contractions.
      The two circles connecting the left and right operators 
      ${\mathbb U}$ and ${\mathbb V}$ of the left and right sectors
      stand for diagonal matrices corresponding to the singular values of the two SVD,
      thus implying $j = j'$ and $l = l'$.
      Expliciting the index mapping which leads to the eigenvalue Eq.~(\ref{eq:GenEV}),
      we get the following component-by-component equation:
      $[\HamEff_k]^{i'_k,i_k}_{\lambda \rho, \mu \nu} \, [x_k]^{i_k}_{\mu \nu} = 
      \lambda \, [\Norm_k]^{i'_k,i_k}_{\lambda \rho, \mu \nu} \, [x_k]^{i_k}_{\mu \nu}$,
      where the eigenvector corresponding to minimum eigenvalue
      $[\bar{x}_k]^{i_k}_{\mu \nu} := (A^{[k],i_k})_{\mu,\nu}$.
}
    \label{fig:diagonal}
  \end{center}
\end{figure}

Here we stress that, for each optimization step, once all the required products 
of transfer matrices are constructed, 
of course operators relative to the three sectors have to be contracted in a circle
(see lines connecting blocks in Fig.~\ref{fig:diagonal}) so to build up 
the effective Hamiltonian $\HamEff_k$ and the effective norm operator $\Norm_k$ for a given site $k$.
This operation can be performed fast, due to the truncation in the SVD of the left and right sectors.
Then one proceeds with the stabilization of the generalized eigenvalue problem~(\ref{eq:GenEV}),
following Sec.~\ref{sec:stabilization}, before solving it.
Its solution eventually provides the optimized entries of the matrices $A^{[k]}$
in the MPS state at position $k$.

In the following we will use for the first time the proposed MPS algorithm for PBC
in order to compute the energy response in a spin ring, when subjected
to a change in the boundary conditions (twisted boundary conditions). 
This permits to quantify the {\it spin stiffness} (or analogously the superfluid density in bosonic models).
Hereafter we will only be interested in the ground state energies, while expectation values of
operators will not be taken into account. This will enable us to obtain reliable results 
also using MPS descriptions with a relatively small bond link $m \sim 20$.

\section{Spin stiffness in the Heisenberg chain} \label{sec:Heisenberg}

We consider here the spin-$1/2$ Heisenberg model, that is described by the Hamiltonian:
\beq
\Ham = J \sum_j \left[ \frac{1}{2} \left( S^+_j S^-_{j+1} + {\it h.c.} \right) + \Delta S^z_j S^z_{j+1} \right] \, ,
\label{eq:HeisHam}
\eeq
where $\sigma^\alpha_j = 2 S^\alpha_j$ ($\alpha =x,y,z$) denote the spin-$1/2$ Pauli matrices on site $j$,
$S^{\pm} = S^x \pm i S^y$ are the raising/lowering spin operators,
$J$ is the coupling strength and $\Delta$ is the anisotropy.
Hereafter we will set $J = 1$ as energy scale, and use units of $\hbar = k_b = 1$.

In the thermodynamic limit and at zero temperature, this model exhibits 
a gapless critical phase for $\vert \Delta \vert \leq 1$ 
with quasi-long-range order emerging from power-law decaying correlation functions, 
while it is gapped and short-ranged for $\vert \Delta \vert > 1$.
The gapless phase is characterized by ballistic transport, which corresponds to a finite
spin stiffness $\rho_s(\Delta)$ at the thermodynamic limit~\cite{shastry1990}.
This quantity expresses the sensitiveness to a magnetic flux added in the system when PBC
are assumed, and can be formally treated by performing a twist in the boundary conditions.
Namely, the addition of a flux $\phi$ along the $z$ direction corresponds to take the
twisted boundary conditions
\beq
S^{\pm}_{N+1} = S^{\pm}_1 e^{\pm i \phi} , \qquad S^z_{N+1} = S^z_1 \, ,
\label{eq:twist}
\eeq
where $N$ denotes the length of the ring.
The stiffness is then defined as follows:
\beq
\rho_s := N \frac{\partial^2 E_0(\phi)}{\partial \phi^2} \Big\vert_{\phi = 0} \, ,
\label{eq:stiff}
\eeq
where $E_0(\phi)$ is the ground state energy with a magnetic flux of intensity $\phi$.
Here we remark that $\rho_s$ evaluated with OBC is strictly zero, since with open boundary
geometry one can always cancel the effect of the twist by applying suitable gauge conditions
to the spins.

\begin{figure}
  \begin{center}
    \includegraphics[width=13cm]{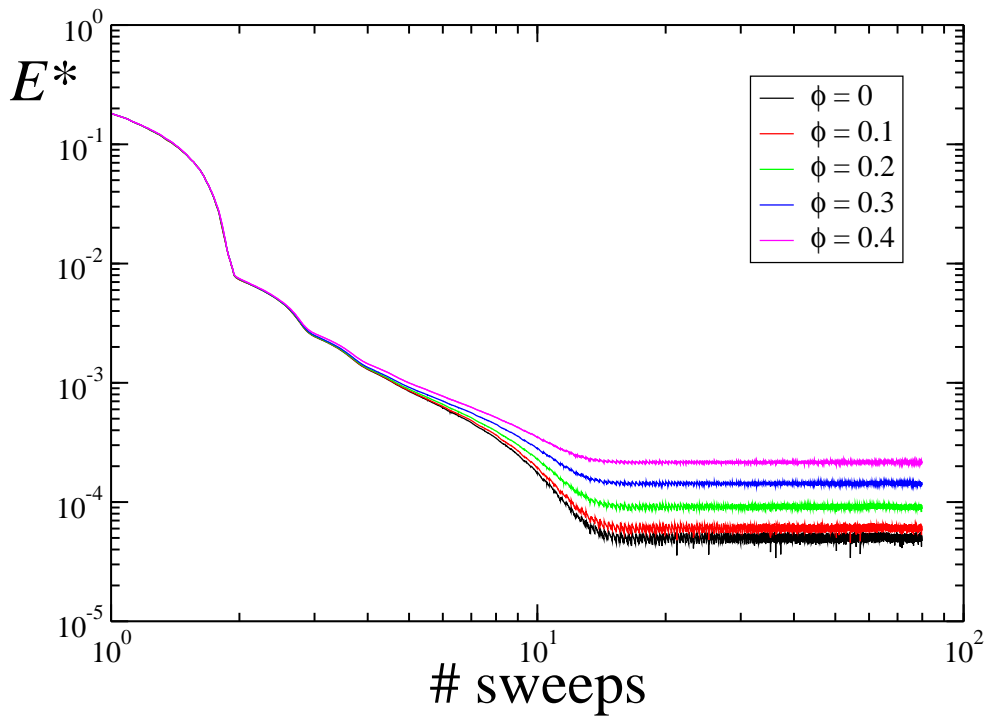}
    \caption{Convergence to the ground-state energy of a spin-$1/2$ Heisenberg chain 
      of length $N=150$ and anisotropy $\Delta = 0.5$ with twisted boundary conditions, 
      as a function of the variational steps in the MPS optimization algorithm.
      Here we used a bond link $m=18$, with truncation indexes $p=50$ and $s=35$;
      we imposed a random initial guess and performed $80$ sweeps
      (energies are rescaled over the ground state energy $E_0$
      with PBC, so that $E^* = E - E_0 + 5 \times 10^{-5}$. 
      For the sake of clarity, we used a logarithmic scale on the $y$ axis).}
    \label{fig:Energy_Heis}
  \end{center}
\end{figure}

By solving the Bethe ansatz equations for model~(\ref{eq:HeisHam}), one can show~\cite{shastry1990} that,
at the thermodynamic limit and for $\vert \Delta \vert \leq 1$,
\beq
\rho_s(\Delta) = J \frac{\pi \sin(\mu)}{4 \mu (\pi - \mu)} \qquad {\rm with} \quad \Delta = \cos(\mu) \, ,
\label{eq:rho_BA}
\eeq
while $\rho_s(\Delta)$ vanishes for $\vert \Delta \vert > 1$.
The point $\vert \Delta \vert = 1$ is characterized by a metal-insulator transition between 
a metallic phase with a finite $\rho_s$ and a gapped insulating regime with $\rho_S=0$,
following a Mott mechanism.

\begin{figure}
  \begin{center}
    \includegraphics[width=12cm]{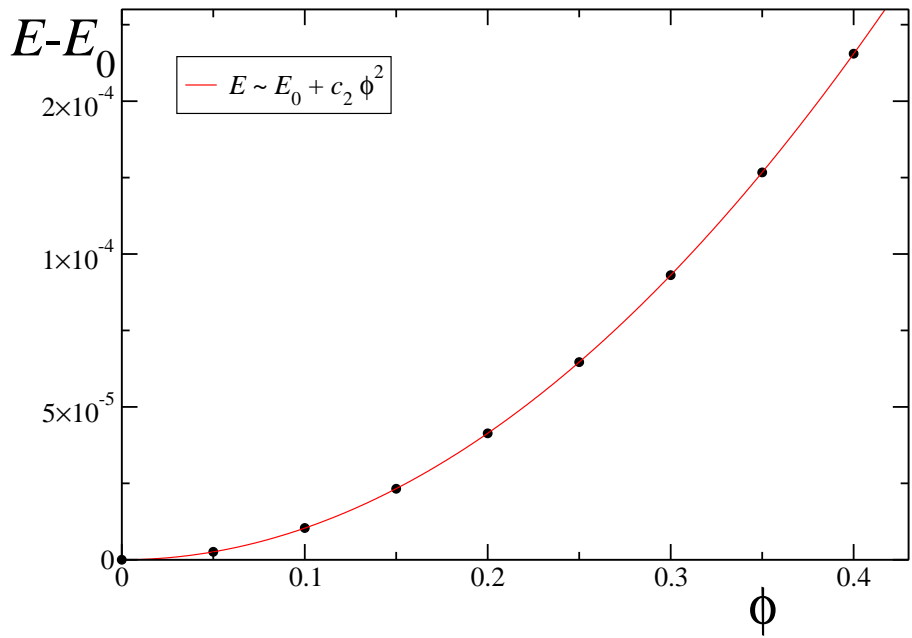}
    \caption{Ground-state energy of a Heisenberg chain of length $N=150$, $\Delta = 0.5$,
      as a function of the twist $\phi$, obtained
      by averaging the data in Fig.~\ref{fig:Energy_Heis} over the last five sweeps.
      The red curve is a quadratic fit of numerical data (black circles),
      with $c_2 \approx 1.03435 \times 10^{-3}$.}
    \label{fig:Energy_HeisB}
  \end{center}
\end{figure}

We employ the MPS algorithm with PBC described above in order to evaluate the stiffness.
In Fig.~\ref{fig:Energy_Heis} we show the actual system energy at each step of the 
MPS algorithm. As one can notice, since the algorithm is variational 
in the energy of the system, the leading behavior of the curve is monotonic decreasing.
When the energy is close to its convergence point, fluctuations due to the truncation
in the SVD process become evident~\footnote{To be more quantitative on this point, 
we need to address the effects in the convergence of the MPS algorithm with 
the two truncation parameters $p$ and $s$. Typically, increasing the truncations
also increases the fluctuations in the converged energy.
This is quite different from the energy behavior by varying the bond link $m$, 
which strictly governs the accuracy of the algorithm.
We will explain this point more in detail later, in Sec.~\ref{sec:converg}.
}.
Anyway, provided the two truncation parameters $p$ and $s$
are chosen sufficiently large, as it is the case in Fig.~\ref{fig:Energy_Heis},
an average over the last sweeps is in general sufficient to sweep away all the 
unwanted fluctuations. This clearly emerges from the averaged converged values 
of the energy, plotted in Fig.~\ref{fig:Energy_HeisB} as a function of the twist angle $\phi$ 
for twisted boundary conditions. Indeed a neat quadratic behavior is visible.
We then fitted the curve $E_0(\phi)$ with a quadratic law: 
$E_0(\phi) = E_0(0) + c_2 \phi^2$ (see the red line), obtaining the prefactor $c_2$
which is directly related to the stiffness:
\beq
\rho_s = 2 N c_2 \,. \label{eq:stiff_calc}
\eeq
In the figure we used a bond link $m=18$, which is considerably smaller than 
the one used in actual simulations performed for OBC (for well optimized codes, 
$m$ can typically reach values one order of magnitude larger).
Nonetheless we should also point out that typically large numbers of $m$ are required
for evaluating local observables or correlation functions; on the contrary, here
we are only interested in the ground state energies for which large values of $m$
are less crucial.
As an example, in the case of the isotropic Heisenberg chain, with $m=18$
we found $E_0/N \approx -0.443138$ which is still not far from the thermodynamic
limit value obtained from Bethe ansatz calculations $\epsilon = -\ln 2 + 1/4 \approx -0.443147$.

\begin{figure}
  \begin{center}
    \includegraphics[width=13cm]{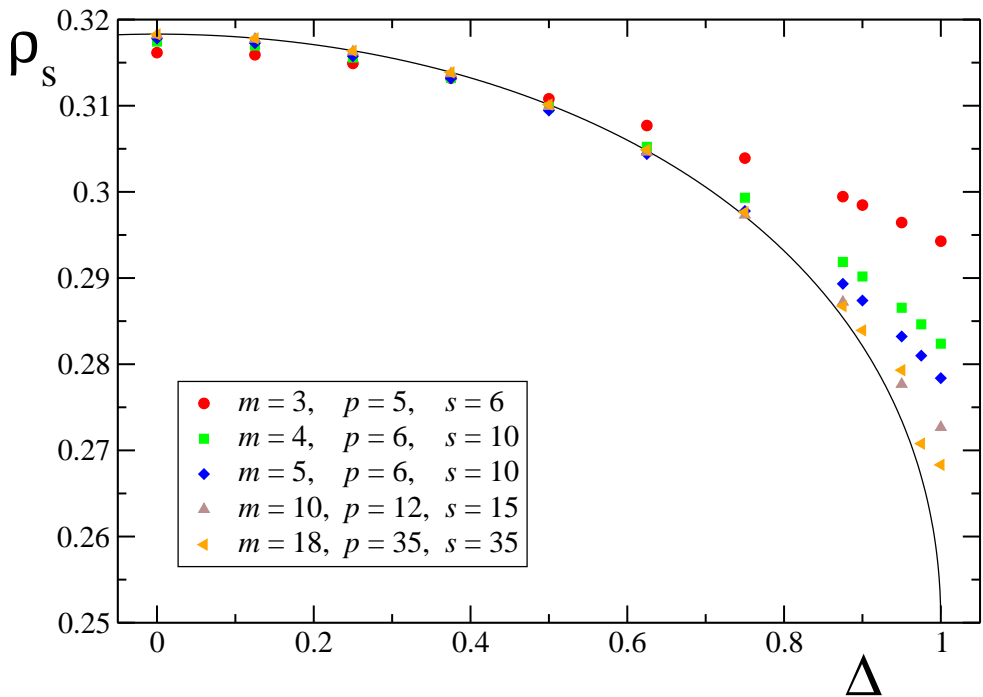}
    \caption{Spin stiffness in the critical phase of the Heisenberg spin-$1/2$ chain,
      Eq.~(\ref{eq:HeisHam}) with $\vert \Delta \vert \leq 1$.
      Symbols are obtained from numerical simulations of a system with $N=180$ sites,
      for different values of the bond link $m$ (and, accordingly, different values of
      $p$ and $s$, see the legend).
      The straight curve is the analytic estimate as obtained from the Bethe ansatz, Eq.~(\ref{eq:rho_BA}).}
    \label{fig:Stiff_Heis}
  \end{center}
\end{figure}

\begin{figure}
  \begin{center}
    \includegraphics[width=12cm]{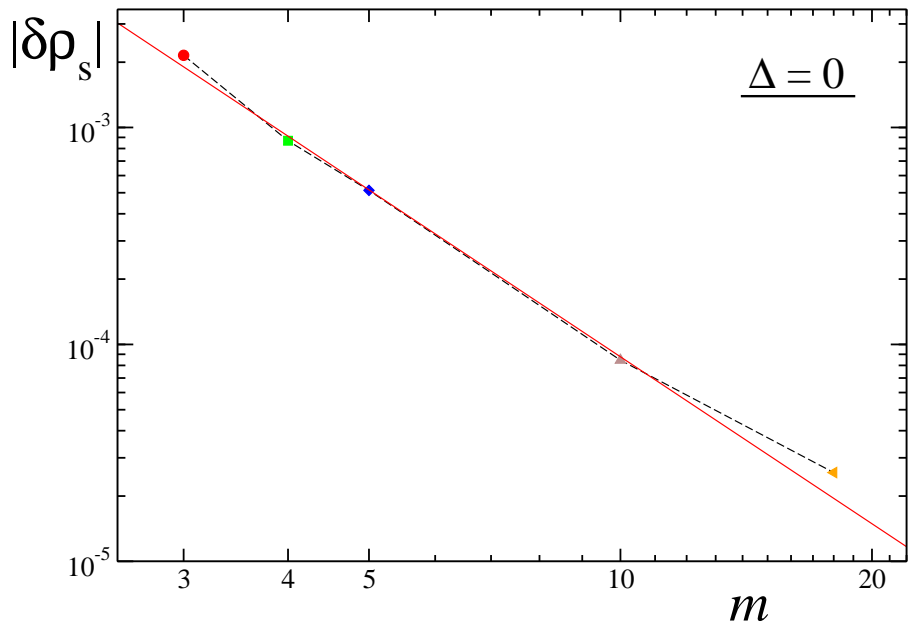}
    \caption{Absolute differences $\vert \delta \rho_s \vert$ between the stiffness in the Heisenberg model 
      with $N=180$ sites, evaluated with a given bond link dimension $m$, 
      and the exact value at the thermodynamic limit, as given by Eq.~(\ref{eq:rho_BA}). 
      The continuous red line denotes a power-law fit of data $\vert \delta \rho_s \vert \sim m^{-2.56}$.
      Data shown are for $\Delta = 0$.}
    \label{fig:Stiff_HeisB}
  \end{center}
\end{figure}

A plot of the spin stiffness~(\ref{eq:stiff}) as a function of the anisotropy $\Delta$,
in the critical phase $\vert \Delta \vert \leq 1$ is
shown in Fig.~\ref{fig:Stiff_Heis}, where data have been collected for systems of $N = 180$ sites. 
We obtained $\rho_S$ as a response in the ground-state energy to a twist in the BCs, 
using Eq.~(\ref{eq:stiff_calc}).
As it is apparent from the figure, modest values of $m$ are sufficient to attain good 
accuracies for $\rho_s$: for values of the anisotropy sufficiently far from the isotropic 
limit, at $m =18$ we are able to reach precisions in $\rho_s$ of the order $\sim O(10^{-5})$.
In particular, the convergence to the exact value at the thermodynamic limit appears
to be approximately power-law in $1/m$, with an exponent $\sim 2.5$ as explicitly shown 
in Fig.~\ref{fig:Stiff_HeisB}.
We also noted that the dependence of the stiffness on the system size $N$ is negligible
on the scale of Fig.~\ref{fig:Stiff_Heis} for $N \gtrsim 10^2$,
even if they can become relevant in a comparison with the exact value 
at the thermodynamic limit, as it is done in Fig.~\ref{fig:Stiff_HeisB}.

Special care has to be taken in the region close to the border of the critical zone 
$\Delta \sim 1$: this requires higher precisions.
Moreover, here the convergence of the minimization algorithm becomes much slower (more than one 
hundred of sweeps are required in order to reach the minimum of energy).
The reason is due to the antiferromagnetic character of the Hamiltonian in this regime~\cite{pirvu2010};
this problem could be at least partly overcome by performing a minimization 
algorithm similar to the one proposed here, but which optimizes two sites at the same time.

\subsection{Convergence with the truncation parameters} \label{sec:converg}

We now discuss the stability of the algorithm with respect 
to the various approximations introduced for speeding up the PBC problem.
Our MPS algorithm for PBC is governed by three control parameters: 
the size $m$ of the matrices in the MPS representation~(\ref{eq:MPS}), the two 
truncation parameters on the singular values of long products 
of transfer matrices ($p$) and on the effective Hamiltonian ($s$).
As we already discussed before and as it is apparent from Fig.~\ref{fig:Stiff_HeisB},
the bond link controls the global accuracy of the algorithm.
This is analogous to the standard DMRG algorithms, where an increase of $m$ 
typically produces a converged value of the ground state energy which monotonically 
decreases towards the exact value.
In our case, we found an approximately power-law convergence with $m$ 
of the stiffness $\rho_s$.

\begin{figure}
  \begin{center}
    \includegraphics[width=13cm]{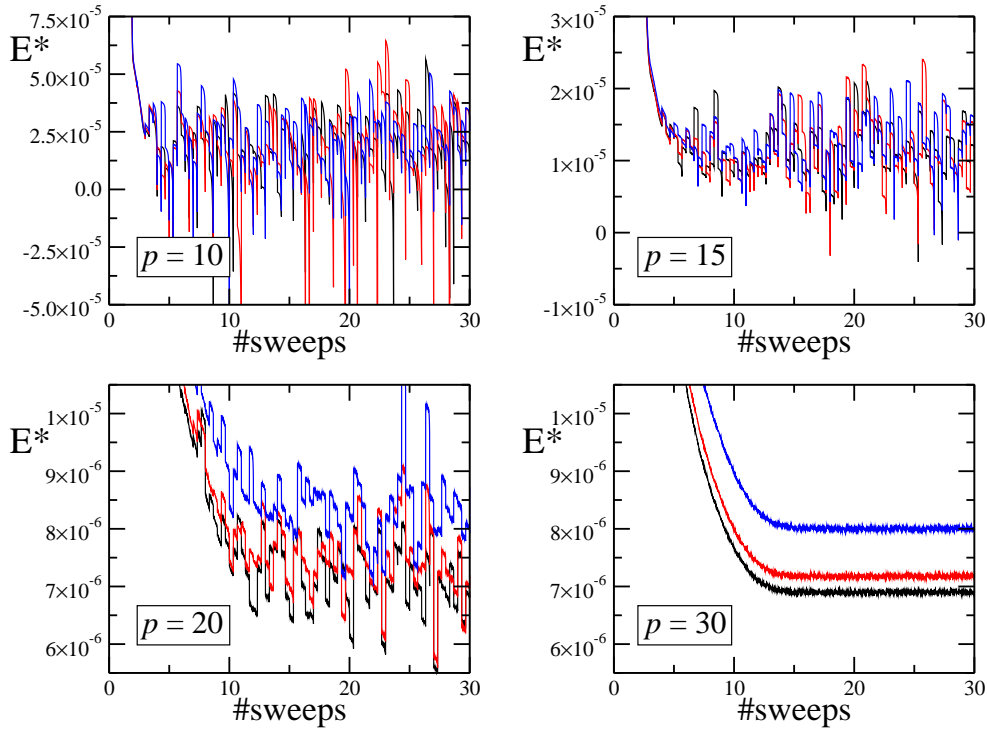}
    \caption{Convergence of the MPS periodic optimization algorithm, as a function of the truncation parameter $p$.
      The different panels display the ground-state energy per-spin of a spin-$1/2$ Heisenberg chain 
      of length $N=150$ and anisotropy $\Delta = 0.5$ with twisted boundary conditions, 
      as a function of the variational steps. Here we fixed the bond link $m=18$, and the index $s=35$
      (energies are rescaled such that $E^* = E + 0.375$).
      The four panels are for $p = 10, \, 15, \, 20, \, 30$.
      Note that the scale on the $y$ axis of the upper left, the upper right, and the two lower panels
      are different.}
    \label{fig:Converg_p}
  \end{center}
\end{figure}

Let us concentrate on the effects of the truncations.
We first discuss the convergence of the algorithm with $p$.
This parameter expresses the number of singular values that are kept in the SVD
representation of the product of transfer matrices belonging to sectors
other than the one on which the optimization is running on (see Sec.~\ref{sec:opt}
and Fig.~\ref{fig:circular}), and it ranges in the interval $[1,m^2]$.
Only in the case $p=m^2$ no errors are introduced in the SVD representation;
any $p<m^2$ will introduce further errors in the algorithm.
As it is clearly visible from Fig.~\ref{fig:Converg_p}, where in the
different panels we show the convergence of the ground state energy
for different values of $p$ fixing $m$ and $s$, a value of $p < m^2$
introduces non-monotonic fluctuations in correspondence of any
change of sector, during the optimization algorithm (therefore
with periodicity $1/3$, in units of the number of sweeps).
On the contrary, while the algorithm is running in a given sector 
without changing it, the energy is monotonically decreasing,
due to the intrinsic variational character of the optimization.
When $p$ is increased, these fluctuations diminish, until they
become hardly visible on the scale of the figure for $p \gtrsim m$
(note that the energy scale in the panels of Fig.~\ref{fig:Converg_p} is different).
For example, in the case $m=18$ we obtained quite stable results for $p \gtrsim 30$.
We remark that for low values of $p$ ($p=10, \, 15$ in the two upper
panels of Fig.~\ref{fig:Converg_p}), due to strong fluctuations, we could not 
even see an increase of the energy with the magnetic flux $\phi$, therefore it
was impossible to extract a value for the spin stiffness $\rho_s$.

\begin{figure}
  \begin{center}
    \includegraphics[width=13cm]{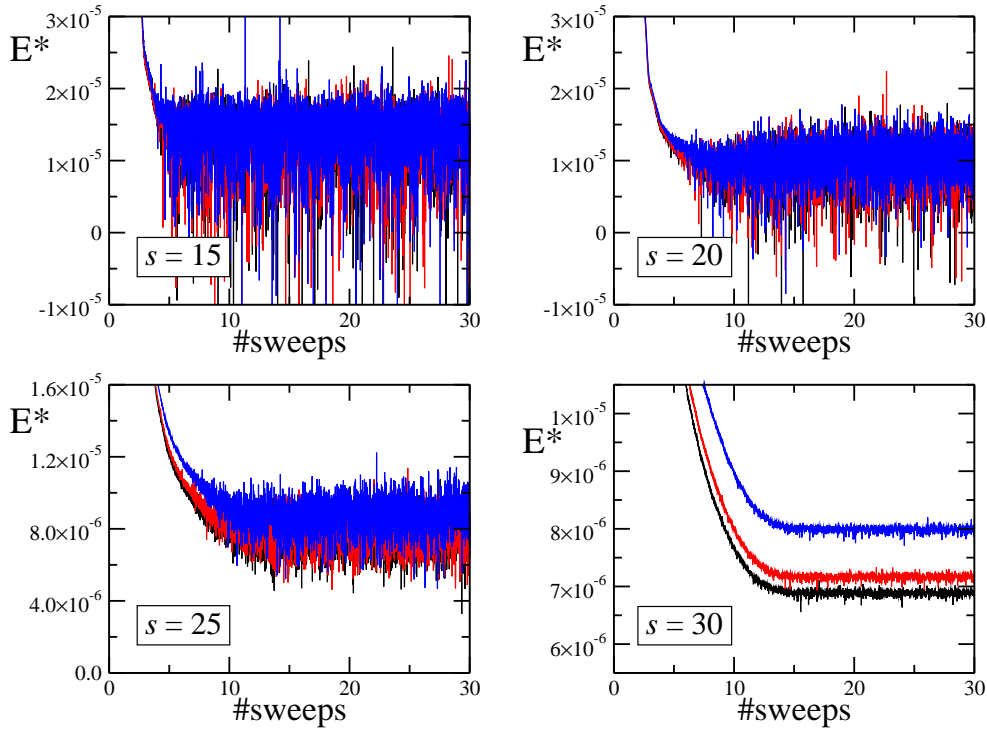}
    \caption{Same as in Fig.~\ref{fig:Converg_p}, but keeping $p=35$ fixed and 
      varying the truncation parameter $s = 15, \, 20, \, 25 \, 30$.
      Note that the scale on the $y$ axis of the two upper panels, the lower left and the lower right panel
      are different.}
    \label{fig:Converg_s}
  \end{center}
\end{figure}

We finally focus on the convergence with $s$, which quantifies the number of singular values 
kept in the SVD of the effective Hamiltonian $\HamEff_k'$.
It turns out that, as it happens with the other truncation parameter $p$, by increasing 
$s \in [1, m^2]$ we found a decrease of the fluctuations (see Fig.~\ref{fig:Converg_s}).
However these fluctuations can be distinguished from the ones due to $p$,
since they do not have a definite periodic structure and are random, 
as a function of the steps in the algorithm.
This is because at each variational step one has to construct an effective Hamiltonian.
Like for $p$, we found that a value $s \gtrsim m$ is able to reduce 
such fluctuations such to obtain stable values for the converged energy
(in the example of Fig.~\ref{fig:Converg_s} with $m=18$, we noted that 
$s \gtrsim 30$ is sufficient to compute $\rho_s$).

We point out that the correct evaluation of the susceptibility with respect to $\phi$ 
in Eq.~(\ref{eq:stiff}) needs an high degree of convergence of the energy $E(\phi)$.
As a matter of fact, fluctuations caused by low values of the truncation parameters
may completely hide the small variations of $E$ induced by infinitesimal magnetic fields $\phi$,
thus invalidating our procedure for the evaluation of $\rho_s$
(as it is visible if the upper panels of Figs.~\ref{fig:Converg_p},~\ref{fig:Converg_s}).
In conclusion, as already stated in Sec.~\ref{sec:SVD}, for our calculations
we required $p,s \gtrsim m$, thus practically worsening the computational requirements
of the PBC algorithm to $O(m^4)$, instead of $O(m^3)$ as it is claimed
in Ref.~\cite{pippan2010}.
On the other hand, in order to give a rather precise estimate of the ground state
energy $E(\phi = 0)$, it is probably sufficient to take lower values of $p$
(as in the upper left panels of Figs.~\ref{fig:Converg_p},~\ref{fig:Converg_s}, 
where the relative error induced by the fluctuations is already
$\delta E / E_0 \sim 10^{-6}$, with $\delta E$ being the size 
of fluctuations and $E_0$ the ground state energy).
This also explains why Pippan {\it et al.} were able to go to larger values 
of $m$ ($m \sim 50$) but apparently smaller values of $p$, in order to get 
the ground state energy of the $s=1$ Heisenberg model with their 
PBC algorithm~\cite{pippan2010}~\footnote{We did not push our simulations 
further in $m$, since already with the parameters used here we obtained 
rather accurate results, apart from the points close to $\Delta = 1$.
However, in a recent paper we were able to reach $m=40$ with reasonable
computational effort, using our MPS periodic algorithm~\cite{spin1-supersolid}.}.

\section{Conclusions} \label{sec:concl}

In this paper we presented a variational procedure for numerically finding the ground 
state of a generically non-translationally invariant one-dimensional Hamiltonian model, 
which can be accurately written as a suitable MPS.
This is constructed starting from the work in Ref.~\cite{pippan2010}, which in turn 
is an improvement of the variational formulation of DMRG given in~\cite{verstraete2004}.
On top of this, we elucidated some technical improvements for stabilizing the code,
as detailed in Sec.~\ref{sec:stabilization}, which also reveal useful in the precision 
measurements of energies that we performed subsequently.

The algorithm globally scales as $O(N \times p \times m^3 )$, where $N$ is the system
size, $m$ is the size of each matrix in the MPS representation, while $p$ is the number 
of singular values that are kept in typical SVD decompositions of transfer matrices 
(generally, for good performances one has to take $p \sim m$).
The accuracy of the algorithm is the same as for the open-boundary case, 
once the dimension $m$ of the matrices is fixed in both cases
(it is therefore also the same as for the original DMRG algorithm, where $m$ 
is the analogous of the number of states kept for describing each block).
We point out that, as typically implemented in good DMRG codes, also in this
procedure it is in principle possible to take advantage of some specific symmetries 
of the Hamiltonian under scrutiny. In particular, Abelian U(1) symmetries can be 
included~\cite{followup}. Reasonably, this would admit a computational speedup of up 
to one order of magnitude, as thoroughly noticed in analogous benchmark 
MPS codes with open boundary conditions.

We showed how to exploit the possibility to change the boundary conditions on the ring, 
in order to evaluate the response to a magnetic flux added in the system.
This provides the so called stiffness, which acts as order parameter for 
the critical phase. As an example, we calculated the spin stiffness
in the spin-$1/2$ Heisenberg chain and compared it with the analytic predictions.
The quantitative analysis of the stiffness, in combination with
a study of the solid ordering in strongly correlated systems through structure
factor measurements, is of crucial importance in the characterization of the 
different states of matter which may arise, including elusive ones, such as
the supersolid phase~\cite{spin1-supersolid}.
It is worth stressing that 
our MPS algorithm inherently accesses very large systems, thus directly addressing 
the system properties in the thermodynamic limit.

\appendix

\section{Working principle of the truncated SVD} \label{appa} 

Consider first the case in which the $m^2 \times m^2$ matrix $M$ admits exactly $r$ 
non zero singular eigenvalues with $r\ll p$.  
This implies that its singular decomposition can be expressed as $M= U DV$ with 
$U\in {\cal M}_{m^2\times r}$, $V\in {\cal M}_{r\times m^2}$ isometries, and 
$D= {\cal M}_{r\times r}$ diagonal. 
Now observe that, given $z$ a vector of $\mathbb{C}^{r}$, $Uz$ is a vector of $\mathbb{C}^{m^2}$. 
Define then the subspace $\mathfrak{A} \subset \mathbb{C}^{m^2}$ spanned by  these vectors, i.e., 
\begin{eqnarray}
  \mathfrak{A} = \mbox{Span}\{ Uz : z \in \mathbb{C}^{r}\} \;.
\end{eqnarray} 
By construction it has dimension $r$ and its orthogonal complement is the left-kernel of $M$ 
(i.e., it is formed by the  vectors  of  $\mathbb{C}^{m^2}$  which nullify
when we apply $M$ on their right).  
Let us then consider a full rank random matrix $x\in {\cal M}_{p \times m^2}$ with $p< m^2$: 
by construction its rows $x_1, x_2, \cdots, x_p$ will span a $p$-dimensional subspace $\mathfrak{B}$
of $\mathbb{C}^{m^2}$.  Since $r\ll p$, we can assume that $\mathfrak{B}$ will have 
a non trivial overlap with the subspace $\mathfrak{A}$ (i.e., no non trivial subspace 
of the latter will be fully disconnected from the former). 
%
By the same token we also notice that for $\ell=1, \cdots, p$, the vectors $y_\ell = x_\ell M$ 
will in general span a subspace $\mathfrak{C}$ of $\mathbb{C}^{m^2}$ of dimension no larger 
than $r$, which with high probability coincides with the image of  $\mathbb{C}^{m^2}$ 
generated via the application of $M$ on the right. 
Via Gram-Schmidt decomposition we construct now an orthonormal set of row vectors 
$\{ y_\ell' : \ell =1, \cdots, p\}$ which include such space as a proper subspace: using these vectors 
as rows for a $p\times m^2$ matrix, we thus construct the $y'$ matrix of the protocol.
Let  then $w$ be a generic row vector  of $\mathbb{C}^{r}$: by construction $wV$ 
will belong to the $\mathfrak{C}$ space and could be expressed as
$ wV = \sum_{\ell} \alpha_\ell   y_\ell'$. This yields the identity 
$V= V y'^\dag y'$ where $y'$ is the $p \times m^2$ matrix whose rows are given by the vectors $y_\ell'$. 
Now define the $m^2\times p$ matrix $Z= My'^\dag$ and compute its SVD decomposition 
$Z= \tilde{U} \tilde{D} \tilde{V}$: since by construction we have that $Z y' = M y'^\dag y'= M$ 
we can conclude that $M= \tilde{U} \tilde{D} \tilde{V}y'$ which allows 
us to identify $\tilde{D}$ with $D$, $\tilde{U}$ with $U$ and $\tilde{V}y'$ with the isometry $V$.    

Now consider the case in which $M$ possess $r \ll p$ dominant non zero singular eigenvalues, 
plus others which are negligible (i.e., they are not null as in the previous case, but can still 
be neglected when compared with the first $d$ ones). The previous derivation still holds 
in this case: the main difference being that the approximated eigenvalues obtained 
from $Z$ will correspond to the real ones with an error which scales as $\epsilon$ 
(the latter being the relative magnitude of the small singular eigenvalue when compared 
with the large ones). In this respect, it is worth noticing that the procedure does not 
really produce the first $p$ largest singular eigenvalues of $M$: 
however, admitting that $r\ll p$, it will approximately produce the first $r$ largest ones.

\section*{Acknowledgments}

We thank S. Peotta, P. Pippan and P. Silvi for fruitful discussions. 
D.R. acknowledges support from EU through the project SOLID, under the
grant agreement No. 248629. The authors also acknowledge support from the 
Italian MIUR under the FIRB IDEAS project RBID08B3FM.



\section*{References}

\begin{thebibliography}{99}

\bibitem{bloch2008}
  Bloch I, Dalibard J and Zwerger W,
  2008 {\it Rev. Mod. Phys.} {\bf 80} 885

\bibitem{greiner2002}
  Greiner M, Mandel O, Esslinger T, Hansch T W and Bloch I,
  2002 {\it Nature} {\bf 415} 39

\bibitem{kim2004}
  Kim E and Chan M H W,
  2004 {\it Nature} {\bf 427} 225

\bibitem{white1992}
  White S R,
  1992 {\it Phys. Rev. Lett.} {\bf 69} 2863;
  1993 {\it Phys. Rev. B} {\bf 48} 10345

\bibitem{schollwockRMP}
  Schollw\"ock U,
  2005 {\it Rev. Mod. Phys.} {\bf 77} 259

\bibitem{ostlund1995}
  \"Ostlund S and Rommer S,
  1995 {\it Phys. Rev. Lett.} {\bf 75} 3537

\bibitem{schollwock2011}
  Schollw\"ock U,
  2011 {\it Ann. Phys.} {\bf 326} 96

\bibitem{scholl99}
  Rapsch S, Schollw\"ock U and Zwerger W,
  1999 {\it Europhys. Lett.} {\bf 46} 559

\bibitem{verstraete2004}
  Verstraete F, Porras D and Cirac I J,
  2004 {\it Phys. Rev. Lett.} {\bf 93} 227205

\bibitem{sandvik2007}
  Sandvik A W and Vidal G,
  2007 {\it Phys. Rev. Lett.} {\bf 99} 220602

\bibitem{shi2009}
  Shi Q-Q and Zhou H-Q,
  2009 {\it J. Phys. A: Math. Theor.} {\bf 42} 272002

\bibitem{pirvu2010}
  Pirvu B, Verstraete F and Vidal G,
  2011 {\it Phys. Rev. B} {\bf 83}, 125104

\bibitem{pippan2010}
  Pippan P, White S R and Evertz H G,
  2010 {\it Phys. Rev. B} {\bf 81} 081103(R)

\bibitem{murg2008}
  Verstraete F, Murg V and Cirac J I,
  2008 {\it Adv. Phys.} {\bf 57} 143

\bibitem{cirac2009}
  Cirac J I and Verstraete F,
  2009 {\it J. Phys. A: Math. Theor.} {\bf 42} 504004

\bibitem{eisert2010}
  Eisert J, Cramer M and Plenio M B,
  2010 {\it Rev. Mod. Phys.} {\bf 82} 277

\bibitem{white2005}
  White S R,
  2005 {\it Phys. Rev. B} {\bf 72} 180403(R)

\bibitem{shastry1990}
  Shastry B S and Sutherland B,
  1990 {\it Phys. Rev. Lett.} {\bf 65} 243

\bibitem{followup}
  Silvi P {\it et al.}, in preparation.

\bibitem{spin1-supersolid}
  Rossini D, Giovannetti V and Fazio R,
  2011 {\it Phys. Rev. B} {\bf 83} 140411(R)

\end{thebibliography}
\end{document}